# Large scale calculations of thermoelectric transport coefficients: a case study of γ-graphyne with point defects

Jinghua Liang, Huijun Liu[*], Dengdong Fan, Peiheng Jiang

*Key Laboratory of Artificial Micro- and Nano-Structures of Ministry of Education and School of Physics and Technology, Wuhan University, Wuhan 430072, China*

Defects such as vacancies and impurities could have profound effects on the transport properties of thermoelectric materials. However, it is usually quite difficult to directly calculate the thermoelectric properties of defect-containing systems via first-principles method since very large supercell is required. In this work, based on the linear response theory and the kernel polynomial method, we present an efficient approach that can help to calculate the thermoelectric transport coefficients of a large system containing millions of atoms at arbitrary chemical potential and temperature. As a prototype example, we consider dilute vacancies and hydrogen impurities in a large scale γ-graphyne sheet and discuss their effects on the thermoelectric transport properties.

## I. Introduction

With increasing challenges from energy crisis and global warming, thermoelectric materials have attracted extensive attention since it can directly convert waste heat into electrical energy. A critical goal in the thermoelectric community is to enhance the energy conversion efficiency of a thermoelectric material, which is determined by the dimensionless figure-of-merit $ZT = S^2\sigma T/(\kappa_e + \kappa_L)$. Here $T$, $S$, $\sigma$, $\kappa_e$ and $\kappa_L$ are the absolute temperature, the Seebeck coefficient, the electrical conductivity, the electronic and lattice contributions to the thermal conductivity, respectively. To improve the $ZT$ values, doping, alloying and nanostructuring [1, 2, 3, 4] are widely employed in the experiments, which also results in various kinds of defects in the material systems. On the theoretical side, the effects of defects and disorders on the thermoelectric properties were investigated using first-principles methods [5, 6, 7, 8], which are however only limited to small systems with number of atoms less than few hundred. In fact, until now, calculating the thermoelectric properties of realistic samples with randomly distributed defects is still computationally demanding from first-principles, since extremely large supercell is usually required to model the system. It is thus highly desirable to find alternative ways to deal with the thermoelectric properties of large scale systems.

---

[*] To whom correspondence should be addressed. E-mail: phlhj@whu.edu.cn

Concurrently, along with the advances of material fabrication techniques and the desire to realize better thermoelectric performance in the earth-abundant and environment-friendly systems, thermoelectric properties of low-dimensional carbon materials have attracted growing interests in recent years [9, 10, 11]. For example, the star material, graphene, was previously shown to possess high electrical conductivity and large Seebeck coefficient [12, 13]. However, due to its ultrahigh thermal conductivity [14], graphene itself is unfavorable for the thermoelectric application. Alternatively, in our previous work [15, 16], we predicted that some carbon allotropes such as graphyne [17, 18, 19] could be promising candidates of high performance thermoelectric materials. For example, it was found [16] that the γ-graphyne, one of the member of graphyne family, exhibits finite band gap and larger Seebeck coefficient, as well as much smaller thermal conductivity compared with graphene.

As mentioned before, doping is a common way to optimize the thermoelectric properties of materials. Meanwhile, some unconventionally introduced vacancies or impurities are inevitable during the fabrication of thin films [20, 21, 22]. It is thus quite necessary to study the influence of such kind of point defects on the thermoelectric performance of γ-graphyne. In this work, we first propose an efficient real space approach to calculate the transport coefficients of large scale systems, which is based on the linear response theory and the kernel polynomial method (KPM) [23, 24] with tight-binding (TB) model derived from first-principles calculations. As a prototype example, we then consider dilute vacancies and hydrogen impurities in a large scale γ-graphyne sheet and discuss their effects on its thermoelectric transport properties.

The rest of the paper is organized as follows. In Section II, we describe in details the model and the real space method for the calculation of thermoelectric transport coefficients. In Section III, we discuss the effects of vacancies and hydrogen impurities on the thermoelectric properties of a large scale γ-graphyne. A brief conclusion is given in Section IV.

## II. Model and Methods

Within the linear response regime, the electrical current $j_e$ and the thermal current $j_q$ are related to the response functions $L_{\mu\nu}$ as

$$\begin{pmatrix} j_e \\ j_q \end{pmatrix} = \begin{pmatrix} L_{11} & L_{12} \\ L_{21} & L_{22} \end{pmatrix} \begin{pmatrix} E \\ -\nabla T \end{pmatrix}, \tag{1}$$

where $E$ is the electric field, and $\nabla T$ is the temperature gradient. The transport coefficients then read as

$$\sigma = L_{11}, \tag{2}$$

$$S = \frac{L_{12}}{L_{11}}, \tag{3}$$

$$\kappa_e = L_{22} - \frac{L_{21}L_{12}}{L_{11}}. \tag{4}$$

It has been verified that the response functions $L_{\mu\nu}$ are determined by the energy-dependent conductivity $\sigma_0(\varepsilon)$ as [25, 26]

$$L_{11} = \int \sigma_0(\varepsilon) \left( -\frac{\partial f(\mu,T)}{\partial \varepsilon} \right) d\varepsilon, \tag{5}$$

$$L_{12} = \frac{1}{T} L_{21} = \frac{1}{eT} \int (\varepsilon - \mu) \sigma_0(\varepsilon) \left( -\frac{\partial f(\mu,T)}{\partial \varepsilon} \right) d\varepsilon, \tag{6}$$

$$L_{22} = \frac{1}{e^2 T} \int (\varepsilon - \mu)^2 \sigma_0(\varepsilon) \left( -\frac{\partial f(\mu,T)}{\partial \varepsilon} \right) d\varepsilon, \tag{7}$$

where $e < 0$ is the elementary charge of electron, $\varepsilon$ the energy of electron, $\mu$ the chemical potential, and $f(\mu,T)$ the Fermi function. As the chemical potential $\mu$ can be shifted up or down by doping or applying external voltage, we will describe its position with respect to the Fermi level $E_F$ of the pristine system in the following discussions.

In the present calculations, the energy-dependent conductivity $\sigma_0(\varepsilon)$ along $\alpha$ direction is determined by the Kubo-Greenwood formula [27]

$$\sigma_0^\alpha(\varepsilon) = \frac{\pi\hbar e^2}{\Omega} Tr\langle v_\alpha \delta(\varepsilon - H) v_\alpha \delta(\varepsilon - H)\rangle, \tag{8}$$

where $\hbar$ is the reduced Planck constant, $\Omega$ the volume of the sample, and $H$ the given Hamiltonian. $v_\alpha = [r_\alpha, H]/i\hbar$ is the $\alpha$ component of the velocity operator. Here we calculate the energy-dependent conductivity $\sigma_0(\varepsilon)$ using an efficient real space method, which is suitable for large scale systems and implemented by expanding the delta function $\delta(\varepsilon - H)$ in Eq. (8) using the KPM [23, 24]. Before the expansion, we need to rescale the Hamiltonian $H$ so that its upper $E^+$ and lower $E^-$ bounds of eigenenergies are mapped to 1 and −1, respectively. We denote the rescaled Hamiltonian and energy by $\tilde{H}$ and $\tilde{\varepsilon}$, and then expand the rescaled delta function in terms of the Chebyshev polynomials as

$$\delta(\tilde{\varepsilon} - \tilde{H}) = \frac{2}{\pi\sqrt{1-\tilde{\varepsilon}^2}} \sum_{m=0}^{M} g_m \frac{T_m(\tilde{\varepsilon})}{\delta_{m,0}+1} T_m(\tilde{H}), \tag{9}$$

where $T_m(x) = \cos[m\arccos(x)]$ is the $m$-th order Chebyshev polynomial of the first kind, $g_m$ is the Jackson kernel introduced to smooth the Gibbs oscillation induced by the truncation of expansion with a finite number of terms. Finally, the energy-dependent conductivity $\sigma_0(\varepsilon)$ can be expressed as

$$\sigma_0^\alpha(\varepsilon) = \frac{4}{\Delta E^2} \frac{4\hbar e^2}{\pi(1-\tilde{\varepsilon}^2)\Omega} \sum_{m=0}^{M}\sum_{n=0}^{M} \mu_{mn}^\alpha T_m(\tilde{\varepsilon}) T_n(\tilde{\varepsilon}), \tag{10}$$

where

$$\mu_{mn}^{\alpha} = \frac{g_m g_n}{(\delta_{m,0}+1)(\delta_{n,0}+1)} Tr\langle v_\alpha T_m(\tilde{H}) v_\alpha T_n(\tilde{H})\rangle, \qquad (11)$$

is the energy-independent moment, and $\Delta E = E^+ - E^-$. Note that, from Eq. (9), we can also obtain the density of states (DOS)

$$\rho(\varepsilon) = \frac{2}{\Delta E} \frac{2}{\pi\sqrt{1-\tilde{\varepsilon}^2}} \sum_{m=0}^{M} g_m \frac{T_m(\tilde{\varepsilon})}{\delta_{m,0}+1} Tr\langle T_m(\tilde{H})\rangle. \qquad (12)$$

The most time-consuming part of our calculations resides in the evaluation of moments $\mu_{mn}^{\alpha}$. For a large scale system, instead of full calculation of the trace in Eq. (11), one can replace it by the average of a small number of random phase states [28]. Once the moments $\mu_{mn}^{\alpha}$ are obtained, we can calculate the energy-dependent conductivity $\sigma_0(\varepsilon)$ by Eq. (10), and then evaluate the thermoelectric transport coefficients at arbitrary chemical potential $\mu$ and temperature $T$ through Eqs. (2) ~(7).

As a prototype example, we apply the methods presented above to investigate the effects of vacancies and hydrogen impurities on the thermoelectric properties of γ-graphyne. As shown in Fig. (1), the γ-graphyne comprises 12 carbon atoms in a primitive cell and can be viewed as modified graphene by periodically inserting C(*sp*)-C(*sp*) bonds into some selected C(*sp*$^2$)-C(*sp*$^2$) bonds in the honeycomb lattice.

The TB Hamiltonian of γ-graphyne can be written as

$$H = H_0 + H_{imp}, \qquad (13)$$

where $H_0$ represents the interactions of carbon atoms in the γ-graphyne, $H_{imp}$ denotes the interactions between the impurities and γ-graphyne. Liu *et al.* [29] showed that the electronic structure of clean γ-graphyne can be described by a simple π-electron model, which is given by

$$H_0 = \sum_{<i,j>} t_{ij} c_i^+ c_j, \qquad (14)$$

where $t_{ij}$ is the hopping energies between nearest-neighbor carbon atoms. There exists three kinds of C-C bonds in γ-graphyne, i.e., C(*sp*)-C(*sp*), C(*sp*)-C(*sp*$^2$) and C($sp^2$)-C($sp^2$), which correspond to three different nearest-neighbor hopping energies $t_1$, $t_2$ and $t_3$, respectively (labeled in Fig. 1). The impurity term $H_{imp}$ can be expressed as

$$H_{imp} = \sum_i (V_i d_i^+ c_i + h.c.) + \sum_i \varepsilon_i d_i^+ d_i, \quad (15)$$

where $\varepsilon_i$ is the on-site potential of impurities, $V_i$ the hopping parameter between impurities and carbon atoms in γ-graphyne. In the present work, we obtain the TB parameters $t_{ij}$ for γ-graphyne from Ref. [29], and derive $\varepsilon_i$ and $V_i$ for hydrogen impurities from first-principles calculations by fitting the calculated band structures [30].

Following the convention in Ref. [16], we define the volume of γ-graphyne as its area multiplied by a "realistic" thickness of 3.35Å. To model vacancy defect, we can remove carbon atoms by setting the corresponding hopping parameters to any other sites as zero. Here we consider a large scale γ-graphyne sheet containing $N = 12 \times 290 \times 290 \approx 10^6$ carbon sites with a low concentration of randomly distributed vacancies or hydrogen impurities. For convenience, periodic boundary conditions are used in our calculations. The Chebyshev polynomials are expanded up to the order of $M = 5000$. In the actual evaluation of the moment $\mu_{mn}^\alpha$ in Eq. (11), we use $N_S$ random phase states to calculate the trace, and average $\sigma_0(\varepsilon)$ over $N_R = 20$ disorder realizations. It is found that the results converge well when $N_S N_R > 100$. Besides, we adopt our previous first-principles results of lattice thermal conductivity $\kappa_L$ [16] to predict the *ZT* values. For simplicity, we assume $\kappa_L$ does not change significantly by the vacancies and hydrogen impurities since we are dealing with rather small concentration of defects.

### III. Results and Discussion

We first consider vacancies in γ-graphyne sheet. The calculated DOS $\rho(\varepsilon)$ with different vacancy concentrations $n_v$ (= 0 ~ 2%) are shown in Fig. 2. For the pristine γ-graphyne ($n_v$ = 0), the calculated band gap of 0.44 eV is consistent with first-principles results of 0.46 eV [16], which indicates the accuracy of our method. When vacancies are introduced, obvious defect states are created within the band gap as indicated by the sharp peaks around Fermi level in Fig. 2. Such vacancy-induced midgap states are also known to occur in other low-dimensional materials [31, 32]. With the increase of $n_v$, we see that the defect band widens, and the spectral weight is enhanced inside the band gap. Meanwhile, the overall $\rho(\varepsilon)$ of valence and conduction bands is smeared by the vacancies. It should be noted that $\rho(\varepsilon)$ is symmetric about the Fermi level, which implies that even with the presence of vacancies, the electron-hole symmetry of in γ-graphyne is still preserved [29]. Such symmetry will also appear in the chemical potential dependence of the transport coefficients, as indicated in the following discussions.

In Fig. 3, we show the calculated transport coefficients and $ZT$ values of γ-graphyne for a series of vacancy concentrations $n_v$, plotted as a function of chemical potential $\mu$ at room temperature. Results for pristine γ-graphyne are also shown for reference. It is surprising to see there is a great suppression of the transport coefficients and $ZT$ values even when very small concentration of vacancies are presented in γ-graphyne. As shown in Figs. 3(a) and 3(b), when the chemical potential $\mu$ enters into the valence bands ($\mu - E_F < -0.22$ eV) or conduction bands ($\mu - E_F > 0.22$ eV), both the electrical conductivity $\sigma$ and the electronic thermal conductivity $\kappa_e$ decrease as $n_v$ increases. In particular, for all the investigated concentrations of vacancies, $\kappa_e$ is reduced to be smaller than the lattice thermal conductivity ($\kappa_L = 76.4$ W/mK) [16] in the range of chemical potential considered. However, when the chemical potential $\mu$ is inside the band gap, small peaks are developed for both $\sigma$ and $\kappa_e$ (see the insets in Figs. 3(a) and 3(b)) caused by the presence of midgap states. Owing to the electron-hole symmetry, the electrical conductivity $\sigma$ and the electronic thermal conductivity $\kappa_e$ are also symmetric about the Fermi level, which holds for both the pristine and vacancy-contained γ-graphyne.

The Seebeck coefficient $S$ shown in Fig. 3(c) exhibts more complicated chemical potential dependence. For the pristine γ-graphyne, $S$ has two peaks with different signs, which is positive for holes ($\mu - E_F < 0$) and negative for electrons ($\mu - E_F > 0$), as generally found for semiconducting systems. The absolute values of the $S$ peaks can reach 688 μV/K for both types of carriers. With the increasing of vacancy concentration $n_v$, the peak values of $S$ decrease monotonously and the corresponding chemical potentials shift towards the valence or conduction bands. In addition, we detect two secondary peaks with opposite sign to the former ones within the band gap, which is caused by the presence of midgap states as discussed above. Due to the electron-hole symmetry, the relationship $S(\mu - E_F) = -S(E_F - \mu)$ holds such that $S$ vanishes exactly at the Fermi level for the pristine γ-graphyne and those with vacancies.

With the suppression of the thermoelectric transport coefficients, the $ZT$ values of γ-graphyne are greatly reduced by the introduced vacancies even at very small concentration (see Fig. 3(d)). For example, at the lowest investigated vacancy concentration $n_v = 0.25\%$, the maximum $ZT$ value of γ-graphyne is dropped from 0.23 for the pristine system to 0.09 for the one with vacancies. With increasing $n_v$, the $ZT$ values of γ-graphyne will be further decreased.

Adsorbates such as hydrogen impurities are also very common defects in thin films. Following the discussions of γ-graphyne with vacancies, we also calculate the DOS, the transport coefficients, and the $ZT$ values for γ-graphyne with different hydrogen impurity concentrations $n_d$ at room temperature. The results are summarized in Fig.

4 and Fig. 5. Comparing Fig. 4 with Fig. 2, we see that the hydrogen impurities lead to similar effects on the energy spectrum of γ-graphyne as does for the vacancies, e.g., the creation of midgap states and the smearing of $\rho(\varepsilon)$. An important difference is that the $\rho(\varepsilon)$ of γ-graphyne with hydrogen impurities no longer shows electron-hole symmetry with the central peaks of midgap states move slightly away from the Fermi level. This is due to the nonzero on-site potentials $\varepsilon_i$ of the hydrogen impurities. Moreover, the peaks of midgap states created by the hydrogen impurities are not as sharp as those created by the vacancies.

The transport coefficients and *ZT* values of γ-graphyne are also strongly suppressed by the adsorption of hydrogen impurities (see Fig. 5). In fact, at the same defect concentration, such effect of the hydrogen impurities is slightly stronger than that of the vacancies. For instance, at $n = 0.25\%$, the maximum *ZT* value is reduced to 0.06 for holes and to 0.05 for electrons by the hydrogen impurities, while it is decreased to 0.09 for both types of carries by the vacancies. The transport coefficients also manifest the absence of electron-hole symmetry in the γ-graphyne with hydrogen impurities, as clearly indicated in the insets of Figs. 5(a)~(c). Up to now, we have demonstrated that both the randomly distributed vacancies and hydrogen impurities have great effects on the transport coefficients of γ-graphyne and would strongly suppress its thermoelectric performance, even at very small defect concentration. Hence, to optimize the thermoelectric performance of γ-graphyne, it is vitally important to avoid the appearance of such kind of defects and improve the crystallinity during the fabrication process.

Before concluding, we may need to understand the complex behavior of the chemical potential dependence of Seebeck coefficient *S*. From Eqs. (3), (5) and (6), *S* can be viewed as the average value of $(\varepsilon - \mu)$ weighted by $\sigma_0$ [33, 34], which is in turn proportional to the DOS $\rho(\varepsilon)$ and group velocity $v_\alpha$ (see Eq. (8)). Electronic states below or above $\mu$ thus have opposite contribution to *S*. Moreover, due to the presence of factor $-\partial f/\partial \varepsilon$ in the integrals, $S$ is dominated by the states in the energy range of several $k_BT$ around $\mu$. As a result, the absolute value of $S$ would be large if there is a big contrast of $\rho(\varepsilon)$ and/or $v_\alpha$ between states below and above $\mu$. Within this picture, it is easy to understand that, for the pristine γ-graphyne, $S$ would have peaks when $\mu$ is near the band edges where there is a large contrast in $\rho(\varepsilon)$. On the other hand, $S$ is decreased when $\mu$ enters into the valence or conduction bands where the contrast in $\rho(\varepsilon)$ is reduced, and in particular, $S$ become vanishing when $\mu$ moves towards the Fermi level where contributions from holes and electrons compensate each other. When the defects are introduced into γ-graphyne, the peaks of $S$ near band edges would be suppressed since the spectral weight inside the band gap is enhanced and the $\rho(\varepsilon)$ is smeared as we have discussed before. Furthermore, we note that for holes (electrons), the presence of midgap states induces asecondary negative (positive) peak for $S$ (see Figs. 3(c) and 5(c)). This "anomalous" behavior can be explained as follows. Since the midgap states are in higher energy than the valence bands, they contribute different sign to $S$ for hole carriers. Hence, when $\mu$ moves gradually from the valence bands to the Fermi level, the positive $S$ first decreases, then vanishes at certain chemical potential between the valence bands and the midgap states, and finally changes into a negative value. For electron carriers, the variation of $S$ can be understood in the same way as long as we notice that the midgap states are in lower energy than the conduction bands. For the pristine γ-graphyne and those with vacancies, the preserved electron-hole symmetry ensures that the vanishing point is exactly at the Fermi level; while for those with hydrogen impurities that break the symmetry, it will move away from the Fermi level. It should also be mentioned that although the peaks of $\sigma$ and $\kappa_e$ inside the band gap are negligible as compared to those deep inside the valence or conduction bands (see Figs. 3(a)~(b) and 5(a)~(b)), the extra peaks in $S$ created by the midgap states are comparable to those near the band edges, which implies that $S$ is more sensitive to the defects. It is therefore suggested that probing $S$ can serve as a powerful tool to identify the types and concentrations of defects in γ-graphyne, as also previously shown for graphene [35, 36].

**IV. Conclusion**

In summary, based on the linear response theory and the KPM, we have implemented an efficient real space method capable to study the thermoelectric properties of a large scale system. With this approach, we can easily compute the thermoelectric transport coefficients of system containing millions of atoms at arbitrary chemical potential and temperature, once the energy-independent moments are obtained. As a prototype example, we have applied it to investigate the effects of vacancies and hydrogen impurities on the electronic and thermoelectric transport properties of a large scale γ-graphyne sheet. We want to mention that the computational framework presented in this work can be straightforwardly extended to other one-, two- and three-dimensional systems. It is also easy to include the effects of electric [37] or magnetic [23] field by employing the corresponding TB model. We expect that our work may stimulate more theoretical study of the thermoelectric properties of large scale systems, especially when we want to make direct comparison with experimental results where the treatment of small defect concentration are generally beyond the framework of density functional theory (DFT).

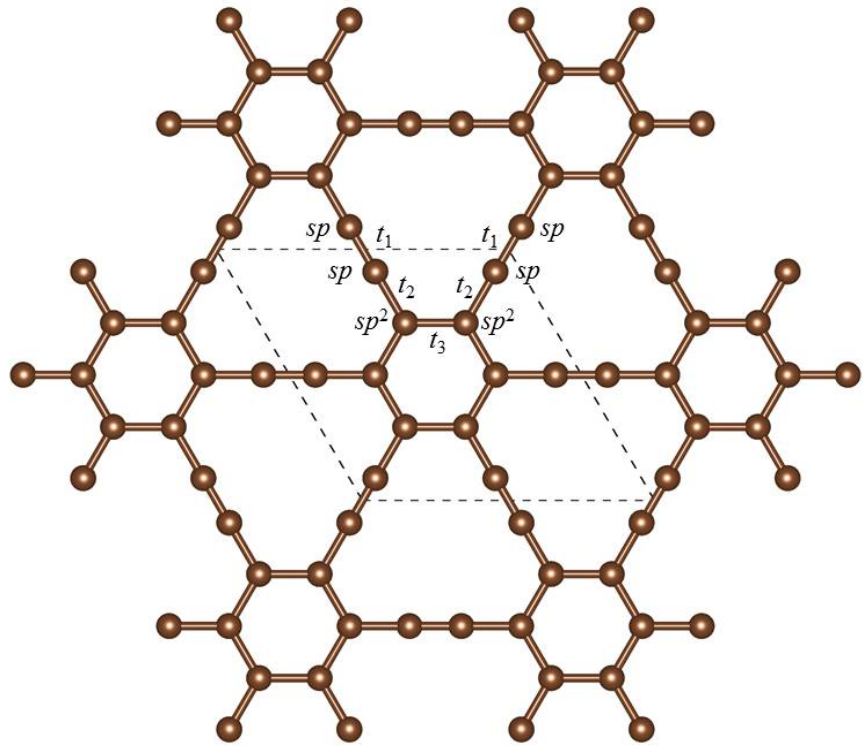

**Figure 1** The ball-and-stick model of γ-graphyne. There are three kinds of C-C bonds, i.e., C(*sp*)-C(*sp*), C(*sp*)-C($sp^2$) and C($sp^2$)-C($sp^2$), which correspond to three different nearest-neighbor hopping energies $t_1$, $t_2$ and $t_3$, respectively. The dashed lines indicate the primitive cell of γ-graphyne.

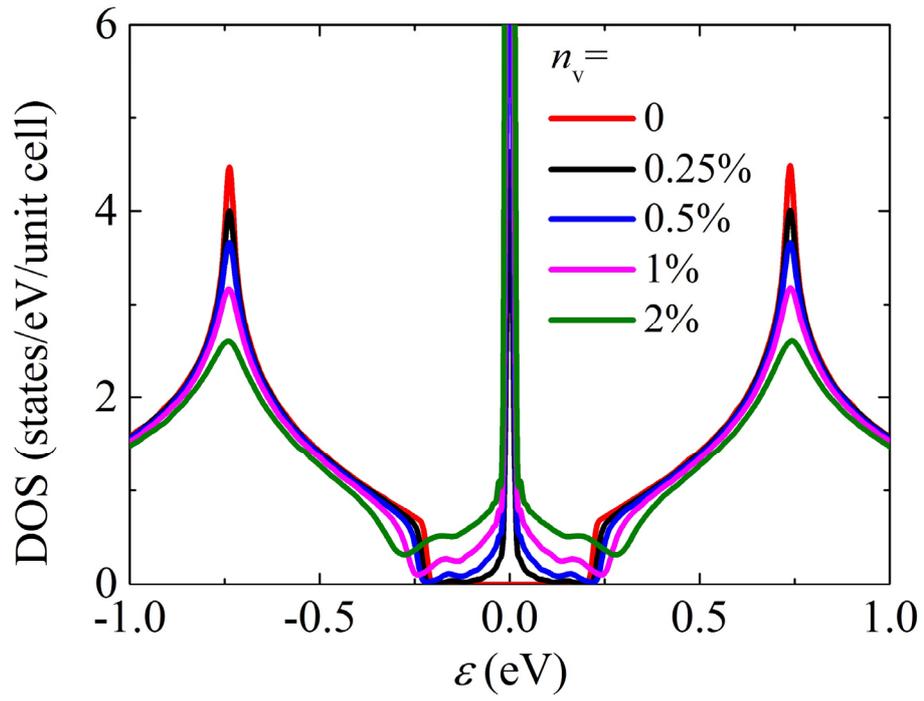

**Figure 2** The calculated DOS $\rho(\varepsilon)$ of γ-graphyne for a series of vacancy concentration $n_v$.

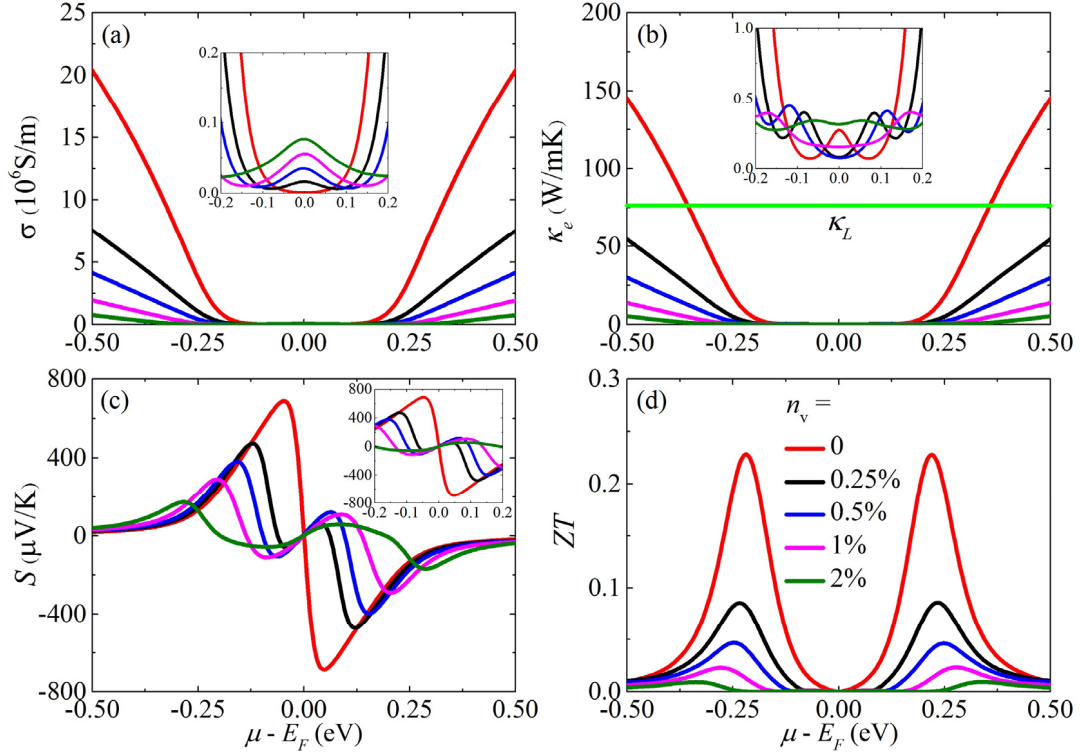

**Figure 3** The calculated (a) electrical conductivity $\sigma$, (b) electronic thermal conductivity $\kappa_e$, (c) Seebeck coefficient $S$, and (d) $ZT$ value of γ-graphyne as a function of chemical potential $\mu$ for a series of vacancy concentration $n_v$ at room temperature. The insets in (a), (b) and (c) show a zoom of transport coefficients inside the band gap. The lattice thermal conductivity $\kappa_L$ adopted from Ref. [16] is also given for reference.

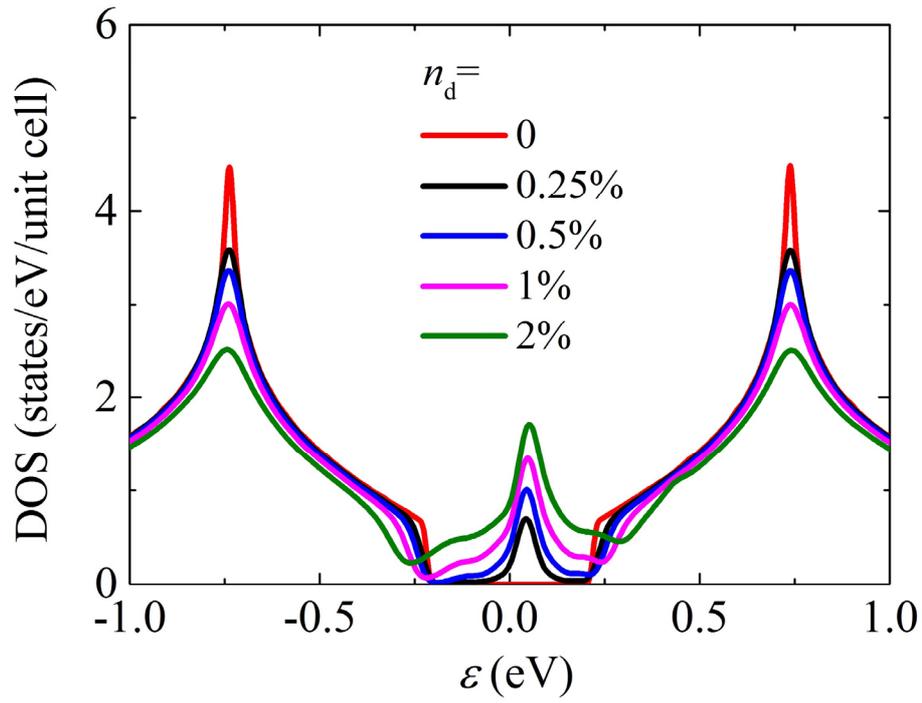

**Figure 4** The calculated DOS $\rho(\varepsilon)$ of γ-graphyne for a series of hydrogen impurity concentration $n_d$.

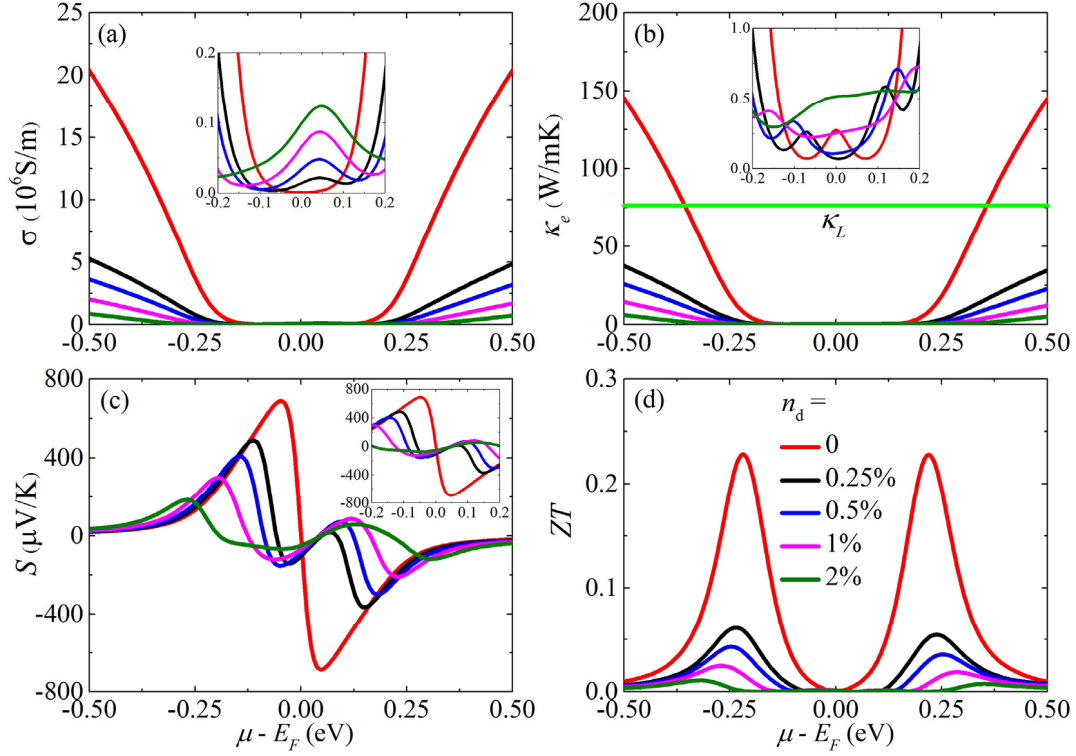

**Figure 5** The calculated (a) electrical conductivity $\sigma$, (b) electronic thermal conductivity $\kappa_e$, (c) Seebeck coefficient $S$, and (d) $ZT$ value of γ-graphyne as a function of chemical potential for a series of hydrogen impurity concentration $n_d$ at room temperature. The insets in (a), (b) and (c) show a zoom of transport coefficients inside the band gap. The lattice thermal conductivity $\kappa_L$ adopted from Ref. [16] is also given for reference.